\documentclass[aps,prl,a4paper,twocolumn,superscriptaddress,citeautoscript,footinbib,floatfix]{revtex4-2}
\usepackage{color}
\usepackage[english]{babel}
\usepackage{lmodern}
\usepackage[T1]{fontenc}
\usepackage{amsmath}
\usepackage{units}
\usepackage{grffile}
\usepackage[bottom]{footmisc}
\usepackage{booktabs}
\usepackage{dcolumn}
\usepackage{graphicx}
\usepackage{lipsum}
\usepackage{changes}
\graphicspath{{figs/}}
\begin{document}

%useful definitions here
\def\EF{$E_\textrm{F}$}
\def\cred{\color{red}}
\def\cblue{\color{blue}}

%
%\title{First principle electronic structure methods
%from pseudohybrid functional with extended Hubbard interactions as an approximation of GW methods}
%
%\title{Emergent Phase Diagram of SrRuO$_{3}$-SrTiO$_{3}$ Heterostructure}
\title{Tunable electronic and magnetic phases in layered ruthenates: \\
SrRuO$_{3}$-SrTiO$_{3}$ heterostructure upon strain}
\author{Minjae Kim}
\email{garix.minjae.kim@gmail.com}\thanks{These two authors contributed equally}
\affiliation{Korea Institute for Advanced Study, Seoul 02455, South Korea}
\author{Chang-Jong Kang}
\email{cjkang87@cnu.ac.kr}\thanks{These two authors contributed equally}
\affiliation{Department of Physics, Chungnam National University, Daejeon 34134, South Korea}
\author{Jae-Ho Han}
\affiliation{Center for Theoretical Physics of Complex Systems, Institute for Basic Science (IBS), Daejeon 34126, South Korea}
\author{Kyoo Kim}
\affiliation{Korea Atomic Energy Research Institute (KAERI), Daejeon 34057, South Korea}
\author{Bongjae Kim}
\email{bongjae.kim@kunsan.ac.kr}
\affiliation{Department of Physics, Kunsan National University, Gunsan 54150, South Korea}
\date{\today}
\begin{abstract}
Layered ruthenates are a unique class of systems which manifests a variety of electronic and magnetic features emerged from competing energy scales. At the heart of such features lies the multi-orbital physics, especially, the orbital-selective behavior. Here, we propose that the SrRuO$_{3}$-SrTiO$_{3}$ heterostructure is a highly tunable platform to obtain the various emergent properties. Employing the density functional theory plus dynamical mean-field theory, we thoroughly investigate the orbital-dependent physics of the system and identify the competing magnetic fluctuations. We show that the epitaxial strain drives the system towards multi-orbital or orbital selective Mott phases from the Hund metal regime. At the same time, the two different types of static magnetism are stabilized, ferromagnetism and checkerboard antiferromagnetism, from the competition with the spin-density wave instability.
\end{abstract}
\maketitle
{\it Introduction.}--
Various physical phenomena found in the layered ruthenates have attracted great interest from the condensed matter physics community.
One of the representative materials is Sr$_2$RuO$_4$.
It shows strange metallic behavior, interpreted as Hund metal phase, in high temperatures ($\gtrsim 25$K)~\cite{Mravlje2011, georges2013strong, Tamai2019, MKim2018, Zhang2016}, and becomes an unconventional superconducting state in low temperatures ($\lesssim 1.5$K)~\cite{Maeno1994}.
In the metallic phase, it is a paramagnet with various magnetic fluctuations~\cite{Steffens2019}. These magnetic fluctuations are thought to be involved as a pairing mechanism of the superconducting state~\cite{Mazin1999, Steffens2019, Roemer2020}, but the exact form of the order parameter is still under dispute after almost thirty years of its finding~\cite{Mackenzie2017}.
Contrastively, another representative material, Ca$_2$RuO$_4$, shows totally different physical properties despite being isovalent to  Sr$_2$RuO$_4$.
The ground state is a Mott insulator
with a static antiferromagnetic order~\cite{nakatsuji1997ca2ruo4,braden1998crystal}.
In between the two systems, Sr$_{2-x}$Ca$_x$RuO$_4$, one can find unusual structural, electric, and magnetic phases
which includes heavy fermionic phases and crossover of local and itinerant magnetism~\cite{Friedt2001,nakatsuji1997ca2ruo4,Carlo2012,kim2021observation}.

One essential source for such interesting behaviors is the multi-orbital nature of the ruthenates.
The orbital-selective electronic correlations and Hund interactions within $t_{2g}$ manifold are found to be responsible for the emergent properties such as Hund metal and diverse magnetic phases in the ruthenates~\cite{Mravlje2011, Dang2015, strand2019magnetic, Steffens2019, georges2013strong, braden1998crystal, nakatsuji1997ca2ruo4, sutter2017hallmarks, gorelov2010nature, liebsch2007subband}. For example, in Ca$_{1.8}$Sr$_{0.2}$RuO$_{4}$, orbital-selective feature is well-identified both from the angle-resolved photoemission spectroscopy and various theoretical frameworks~\cite{kim2021observation, shimoyamada2009strong, neupane2009observation, sutter2019orbitally}.
This multi-orbital nature makes ruthenates as ideal systems for studying the balance and interplay of multiple competing physical parameters, such as bandwidth, inter- and intra-orbital correlation, crystal field splitting, and spin-orbit coupling~\cite{georges2013strong}. In reality, however, there are not many materials other than aforementioned ones, Sr$_2$RuO$_4$, Ca$_2$RuO$_4$, and in-between (Sr$_{2-x}$Ca$_x$RuO$_4$), to be compared with theoretical models on the $t_{2g}$ manifold.

The recent development of the oxide heteroepitaxy has offered a controllable route to tune the physical parameters of materials. Employing ample substrates, one can delicately grow the oxides under various strains with either compressive or tensile-way, which enables the control of crystal field splitting, hopping anisotropy and strength of the correlation~\cite{Schlom2007,BKim2018}. Especially, for strontium ruthenates, experimental demonstrations are readily made with single-layer ruthenates, which is achieved by sandwiching one layer of SrRuO$_{3}$ between SrTiO$_{3}$ blocks~\cite{wu2020thickness,boschker2019ferromagnetism}.
Previously, within the density functional theory (DFT) framework, some of the authors have demonstrated (SrRuO$_{3}$)$_{1}$-(SrTiO$_{3}$)$_{1}$ heterostructure (SRO-STO) as a possible system to study the superconductivity of Sr$_{2}$RuO$_{4}$ based on the similarity of quasi 1-dimensional $xz/yz$ and quasi 2-dimensional $xy$ Fermi surfaces to those from Sr$_{2}$RuO$_{4}$~\cite{kim2020srruo}.
It is interesting to investigate whether the conclusions are valid beyond the DFT framework known as an incomplete description of many-body correlation effects.
%However, the well-known shortcoming of the DFT, insufficient description of many-body correlation effect, dilutes the interesting conclusions.
Especially for ruthenates, the inclusion of the many-body effect is crucial in describing the key electronic features such as Hund-metal, band-renormalization, and metal-insulator transition ~\cite{Mravlje2011,MKim2018,Zhang2016,ricco2018situ,han2018lattice,deng2019signatures,sutter2017hallmarks}, which, in turn, will change the relative strengths of the different magnetism. Considering the prospect of heteroengineering of ruthenates, there lies a plethora of physics that is yet to be discovered through the state-of-the-art computational approaches. \\

In this Letter, we investigate the electronic and magnetic structures of SRO-STO
within the framework of DFT plus dynamical mean-field theory (DFT+DMFT)~\cite{georges1992hubbard,metzner1989correlated,kotliar2006electronic,georges1996dynamical}
for various epitaxial strain and temperature ranges.
We demonstrate that the epitaxial strains controls the orbital-selective electronic correlation and moves the system toward two distinct Mott phase, multi-orbital and orbital selective ones away from the Hund phase.
Even more interesting point is that the strain can tune the strength of the different type of magnetic instabilities and lead to the stabilization of ferromagnetic (FM) or checkerboard-type antiferromagnetic (AFM) orders over spin-density wave (SDW) one.
Here, for completeness, we have employed two different but representative setups in the DMFT approach, and, we map out the full strain-temperature phase diagram of SRO-STO system.

%%%%%%%%%%%%%%%%%%%%%%%%%%%%%%%%%%%%%%%%%%%%%%%%%%%%%%%%%%%%%%%%
\begin{figure}[t]
\includegraphics[width=\columnwidth]{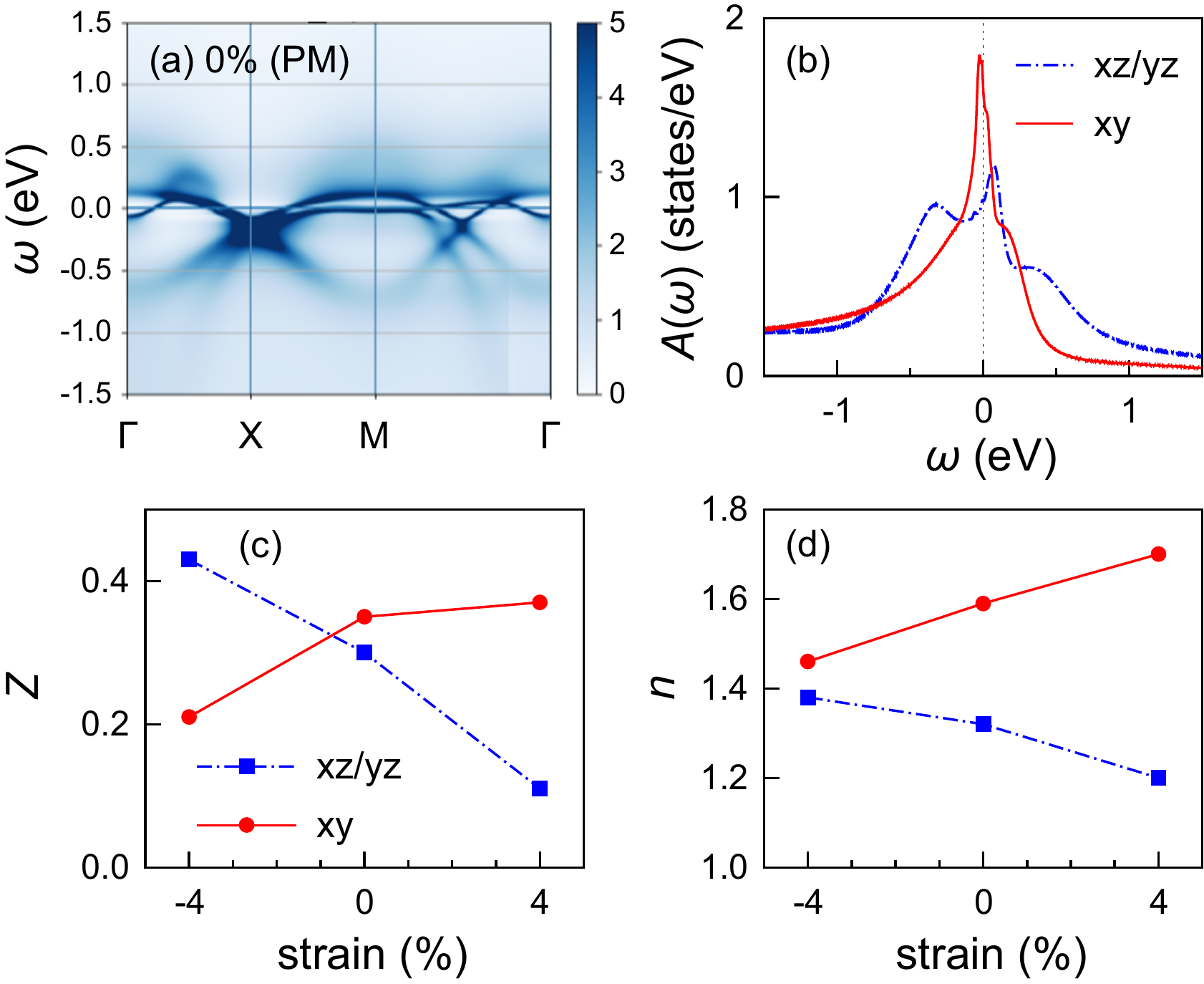}
\caption{(a) The spectral function $A(k,\omega)$
and (b) orbitally resolved density of state $A(\omega)$ of the unstrained SRO-STO (0\% strain). Strain dependent (c) quasiparticle residues, $Z$, and (d) occupancies, $n$, for $xz/yz$ and $xy$ orbitals. The nonmagnetic constraint is forced to simulate the paramagnetism (PM) with the temperature of 35 K for all data.
\label{fig:fig1}
}
\end{figure}
%%%%%%%%%%%%%%%%%%%%%%%%%%%%%%%%%%%%%%%%%%%%%%%%%%%%%%%%%%%%%%%%

{\it Method.}--
We compute the electronic structure of SRO-STO under epitaxial strains of -4\%, 0\%, and +4\% within the DFT+DMFT framework.
We have employed both Ru $t_{2g}$ only and $t_{2g}$+$e_g$ projected basis sets for the DFT+DMFT calculations to check the possible effect of $t_{2g}$-$e_{g}$ hybridization
and confirmed that both give similar results
for spectral functions $A(k, \omega)$ and density of state $A(\omega) = \sum_k A(k,\omega)$ (see more details in supplemental material (SM)~\cite{suppl}).
The DMFT dynamical spin structure factors $S(q,\omega)$ are obtained in the paramagnetic (PM) phase of SRO-STO
with Ru $t_{2g}$+$e_g$ projector scheme~\cite{Park2011}, which covers a wide energy window of [-10, 10] eV from the Fermi level (\EF).
Based on the observation that the spectral functions from the $t_{2g}$ and the $t_{2g}$+$e_g$
projector schemes are consistent (see SM~\cite{suppl}), we present results with the $t_{2g}$ projector scheme unless specified in the text.

{\it Electronic correlation and orbital selectivity.}--
First, we present the electronic structure of the PM state of unstrained SRO-STO (0\% strain) at $T$ = 35 K.
Figures~\ref{fig:fig1} (a) and (b) display related spectral function $A(k,\omega)$ and
its integration over Brillouin zone $A(\omega)$, respectively.
A coherent quasiparticle band is clearly realized near \EF,
which has quasi-two dimensional Ru $xy$-orbital character and exhibits the van Hove singularity (VHS) (see Fig.~\ref{fig:fig1}(b)).
This VHS peak in the unstrained SRO-STO system is much higher than that of Sr$_{2}$RuO$_{4}$~\cite{Fabian2020}.
Hence, enhanced Stoner weight at \EF, and consequently, the stronger tendency towards magnetism is expected for SRO-STO.
In addition, the VHS peak in the SRO-STO system is located just below \EF,
which is the \emph{opposite} to the case of Sr$_{2}$RuO$_{4}$.
Such features could lead to the different electronic and magnetic responses upon external perturbations in the SRO-STO system compared to previously investigated Sr$_{2}$RuO$_{4}$~\cite{Hsu2016,BKim2022}.
The DMFT valence histogram of the unstrained SRO-STO system
exhibits Hund's rule-induced high spin multiplets (see SM~\cite{suppl}),
thereby presenting the Hund metal characteristics.
Especially, we note two archetypical features:
(i) coherence-incoherence crossover as a function of frequency $\omega$ in the self-energy $\Sigma(\omega)$ (see SM~\cite{suppl}),
and (ii) significant electronic correlations in the absence of the Hubbard satellite.
The significant electronic correlations are clearly identified in the quasiparticle residue,
$Z = (1-\partial\Sigma(i\omega)/\partial\omega|_{\omega\rightarrow0^{+}})^{-1}$,
and Ru $xz/yz$ and $xy$ have $Z$ = 0.30 and 0.35, respectively (see Fig.~\ref{fig:fig1}(c)),
which are comparable to the values of Sr$_{2}$RuO$_{4}$, 0.30 ($xz/yz$) and 0.20 ($xy$)~\cite{Fabian2020}.
We see the overall electronic features resemble those for Sr$_{2}$RuO$_{4}$
despite the small differences.

Remarkably, the SRO-STO exhibits diverse electronic phases within an accessible strain range.
The epitaxial strain imposes a tetragonal distortion, and
the energy difference between the Ru $xz/yz$ and $xy$ levels varies under the strain.
Figure~\ref{fig:fig1}(c) displays the strain-dependent quasiparticle residue $Z$ for Ru $xz/yz$ and $xy$ orbitals.
We note the different responses of $Z$ upon tensile strain: decreases for $xz/yz$ orbitals and increases for $xy$ orbital.
This orbital differentiation is clearly reflected in the orbital occupancy as shown in Fig.~\ref{fig:fig1}(d). For the compressive strain, the $xy$ level moves higher in energy and approaches the $xz/yz$ levels (see SM~\cite{suppl}).
As a result, the $xy$-orbital occupation is reduced, and the VHS moves even closer to \EF, indicating stronger FM instability.
These features give rise to the reduction of the low-energy hybridization function for the $xy$ orbital
and, hence, increase the electronic correlations in the $xy$ orbital as shown in Fig.~\ref{fig:fig1}(c).
This can be compared to Ca$_{1.8}$Sr$_{0.2}$RuO$_{4}$~\cite{sutter2019orbitally}, where the occupancy of the $xy$-orbital is close to the half-filling and have a stronger electronic correlation than $xz/yz$ orbitals~\cite{sutter2019orbitally}. For Ca$_{1.8}$Sr$_{0.2}$RuO$_{4}$, the orbital selective Mott-phase is established with gapped $xy$-orbital and metallic $xz/yz$-orbitals~\cite{kim2021observation,shimoyamada2009strong,neupane2009observation,sutter2019orbitally}.
Hence, we argue that the electronic structure of the compressed SRO-STO system corresponds to the $soft$ case of Ca$_{1.8}$Sr$_{0.2}$RuO$_{4}$.

In the case of tensile strains (+4\%), the Ru $xy$ level moves down in energy, and the crystal field splitting between the $xz/yz$ and $xy$ levels becomes larger (see SM~\cite{suppl}). Accordingly, the occupation of the $xz/yz$ orbitals decreases, while that of the $xy$ orbital increases. Contrary to the compressive case, this time, the $Z$ for $xz/yz$ orbitals is 0.11 with the orbital occupation close to the half-filling, 1.20
(see Figs.~\ref{fig:fig1}(c) and (d)).
This means the tensile strain drives the system towards the multi-orbital Mott regime~\cite{suppl,hump}.
The multi-orbital Mott insulating regime is found in Ca$_{2}$RuO$_{4}$,
where the fully-filled $xy$ and half-filled $xz/yz$ orbitals are realized
with a prominent Mott gap of $U+J$ ($\sim$ 2.7 eV)~\cite{sutter2017hallmarks}.

%%%%%%%%%%%%%%%%%%%%%%%%%%%%%%%%%%%%%%%%%%%%%%%%%%%%%%%%%%%%%%%%
\begin{figure}[t]
\includegraphics[width=\columnwidth]{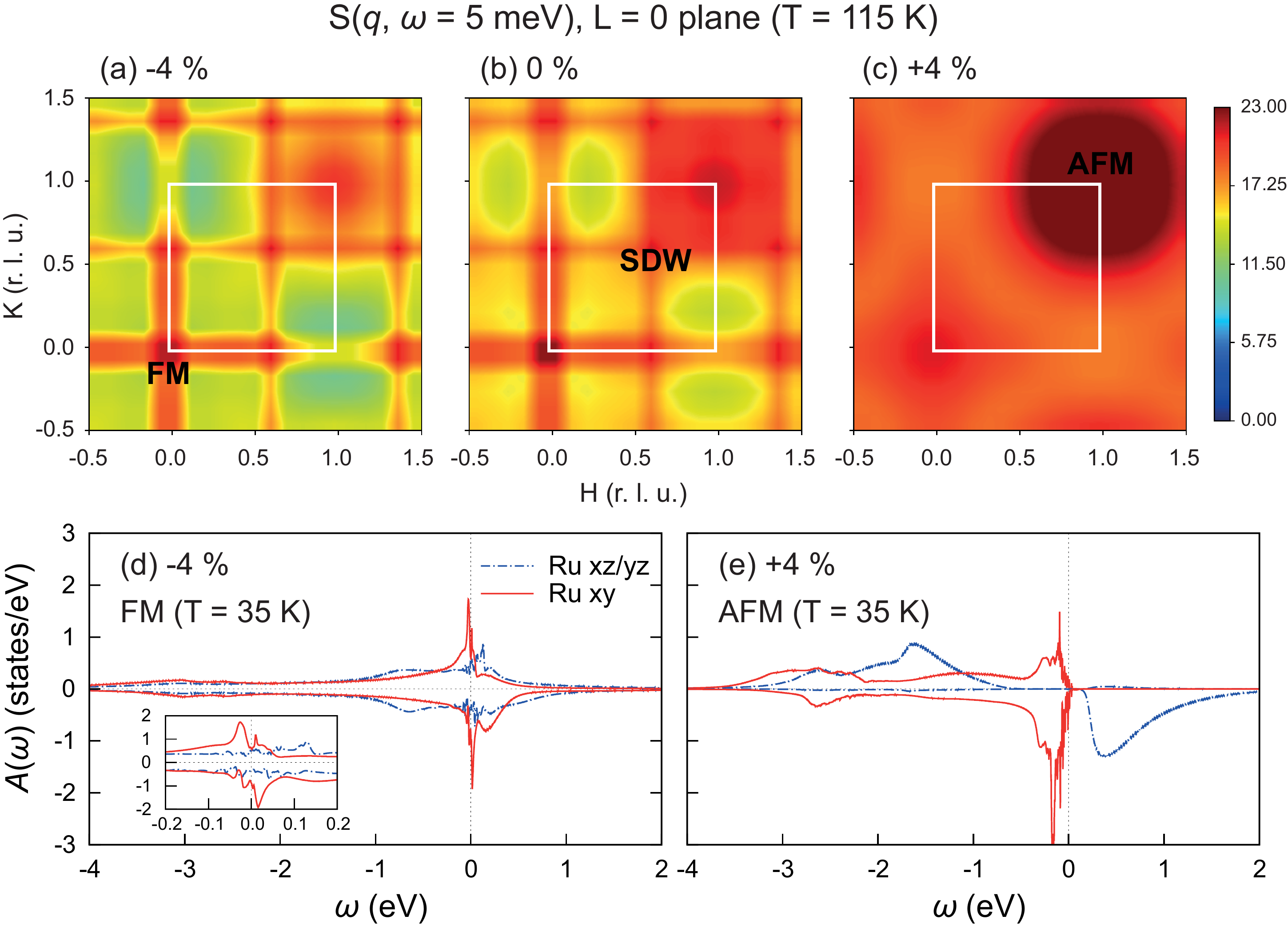}
\caption{Momentum ($q$) dependent dynamical structure factors $S(q,\omega)$ of SRO-STO for (a) -4\%, (b) 0\%, and (c) +4\% strain cases. Here,
for $S(q,\omega)$, the $t_{2g}$+$e_{g}$ projector calculation is performed, and the data is for $\omega$ = 5 meV at the temperature of 115 K. In the lower panels, we present the spin-resolved density of states for (d) FM -4\% strain and (e) AFM +4\% strain cases at 35 K. The inset in (d) is the blowup figure near the Fermi level.
\label{fig:fig2}
}
\end{figure}
%%%%%%%%%%%%%%%%%%%%%%%%%%%%%%%%%%%%%%%%%%%%%%%%%%%%%%%%%%%%%%%%

{\it Magnetism.}--
Now we turn to the magnetic properties of the SRO-STO system. To identify the involved magnetic instabilities, we calculated the DMFT dynamical spin structure factors $S(q,\omega)$ with the local particle-hole vertex correction~\cite{Park2011}. The results are presented in Figs.~\ref{fig:fig2}(a)-(c).
For the unstrained one, three leading competing magnetic instabilities are identified:
FM at $\Gamma$ ($H$, $K$) = (0.0, 0.0), SDW at ($H$, $K$) = ($\sim$ 0.6, $\sim$ 0.6),
and checkerboard AFM at ($H$, $K$) = (1.0, 1.0).
Note that the last instability, AFM, is not captured in the DFT level~\cite{kim2020srruo}, which asserts the importance of the dynamical correlations in describing the magnetism of the system.
Each of the magnetic instabilities is from the different origins: FM from Stoner instability, SDW from Fermi surface nesting of quasi-1D $yz/zx$ character, and AFM from superexchange mechanism~\cite{Mazin1999}. These magnetic instabilities compete without specific dominance and the system remains paramagnetic without static order.
Note that for Sr$_{2}$RuO$_{4}$, two main magnetic instabilities are featured, FM and SDW types, and for Ca$_{2}$RuO$_{4}$, only AFM is stabilized~\cite{strand2019magnetic,Steffens2019,braden1998crystal}.
Hence, we can note that the magnetism involved in the SRO-STO system is even more complex, and also offers
more possibilities for the novel phases include unconventional superconductivity.

For compressive strain (-4\%), the FM instability becomes strongly dominant.
As mentioned before, this is expected as the VHS peak, mostly from $xy$-orbital,
moves closer to~\EF  ~upon the compressive strain (see SM~\cite{suppl})
and the Stoner instability increases accordingly.
As shown in Fig.~\ref{fig:fig2}(a), the AFM instability is strongly suppressed, and the SDW instability has a similar amplitude.
Compared to the unstrained one, the FM tendency in the compressive strain is substantially enhanced and much more predominant concerning the AFM and the SDW.
Hence, the static FM order is set in for the compressive SRO-STO.
We have confirmed that the actual DFT+DMFT calculations converge to the FM ground state under the compressive strain.
The orbital-resolved $A(\omega)$ for the FM ground state is shown in Fig.~\ref{fig:fig2}(d) at $T$ = 35 K.
The exchange splitting (Re$\Sigma_{\downarrow}$(0)-Re$\Sigma_{\uparrow}$(0)) in the Ru $xy$ orbital (199 meV) is larger than that in the Ru $xz/yz$ orbitals (79 meV) (see the inset of Fig.~\ref{fig:fig2}(d)).
This confirms that the Ru $xy$-orbital drives the magnetic transition $via$ the Stoner mechanism.
The non-vanishing magnetic moment started to emerge around $T$ = 100 K
and the value progressively increases as $T$ decreases, thereby suggesting the second-order magnetic transition~\cite{suppl}.
Our obtained magnetic moment is $\sim$ 0.35 $\mu_{B}$, which is close to the experimental report from the ferromagnetic (SrRuO$_{3}$)$_{1}$-(SrTiO$_{3}$)$_{5}$~\cite{boschker2019ferromagnetism}. The saturated moment is much smaller than the
ionic value of 2 $\mu_{B}$, implying the highly itinerant character of the FM.

For the tensile strain of +4\%, the AFM instability strongly prevails and
both the FM and the nesting-induced SDW ones are almost buried under the AFM instability as shown in Fig.~\ref{fig:fig2}(c).
The DMFT density of states $A(\omega)$ at $\omega=0$ (\EF) is reduced from -4\% (compressive) to the 4\% (tensile) strains
(see SM~\cite{suppl}), which manifests the dwindling of the Stoner instability. This is expected
considering the increased interatomic distances between Ru, which reduced the itinerancy of the system.
Besides, we note the SDW Fermi surface nesting is significantly weaker under +4\% strain
(see SM~\cite{suppl}). Such a stark variation of the magnetism upon
the external perturbation is very unique, especially,
in the sense that the other competing magnetic instabilities except for the AFM are almost annulled.
To confirm, we additionally performed the DMFT electronic structure calculations and found that the AFM phase is the ground state under the tensile strain as found from the $S(q,\omega)$ (see SM~\cite{suppl}).
This strong emergence of the checkerboard AFM instability asserts that the mechanism
of the magnetic order now is superexchange interactions between the \emph{local} Ru spins.
Figure~\ref{fig:fig2}(e) shows related orbital-resolved $A(\omega)$ at $T$ = 35 K.
The AFM ground state is an insulating phase with fully occupied Ru $xy$ and half-filled Ru $xz/yz$ orbitals.
The local magnetic moment is about 1.87 $\mu_{B}$/Ru and is remained almost constant below the N\'{e}el temperature of $\sim$100 K,
suggesting the first-order type transition. This is totally different from the FM for compressive strain, where the local moment size develops upon decreasing the temperatures.
We can compare the electronic structure and magnetism of the AFM insulating state of SRO-STO under the tensile strain to those of multi-orbital Mott insulating phase of Ca$_{2}$RuO$_{4}$ at low temperatures,
having a similar electronic configuration, half-filled $xz/yz$ and fully filled $xy$ orbitals.

%%%%%%%%%%%%%%%%%%%%%%%%%%%%%%%%%%%%%%%%%%%%%%%%%%%%%%%%%%%%%%%%
\begin{figure}[t]
\includegraphics[width=\columnwidth]{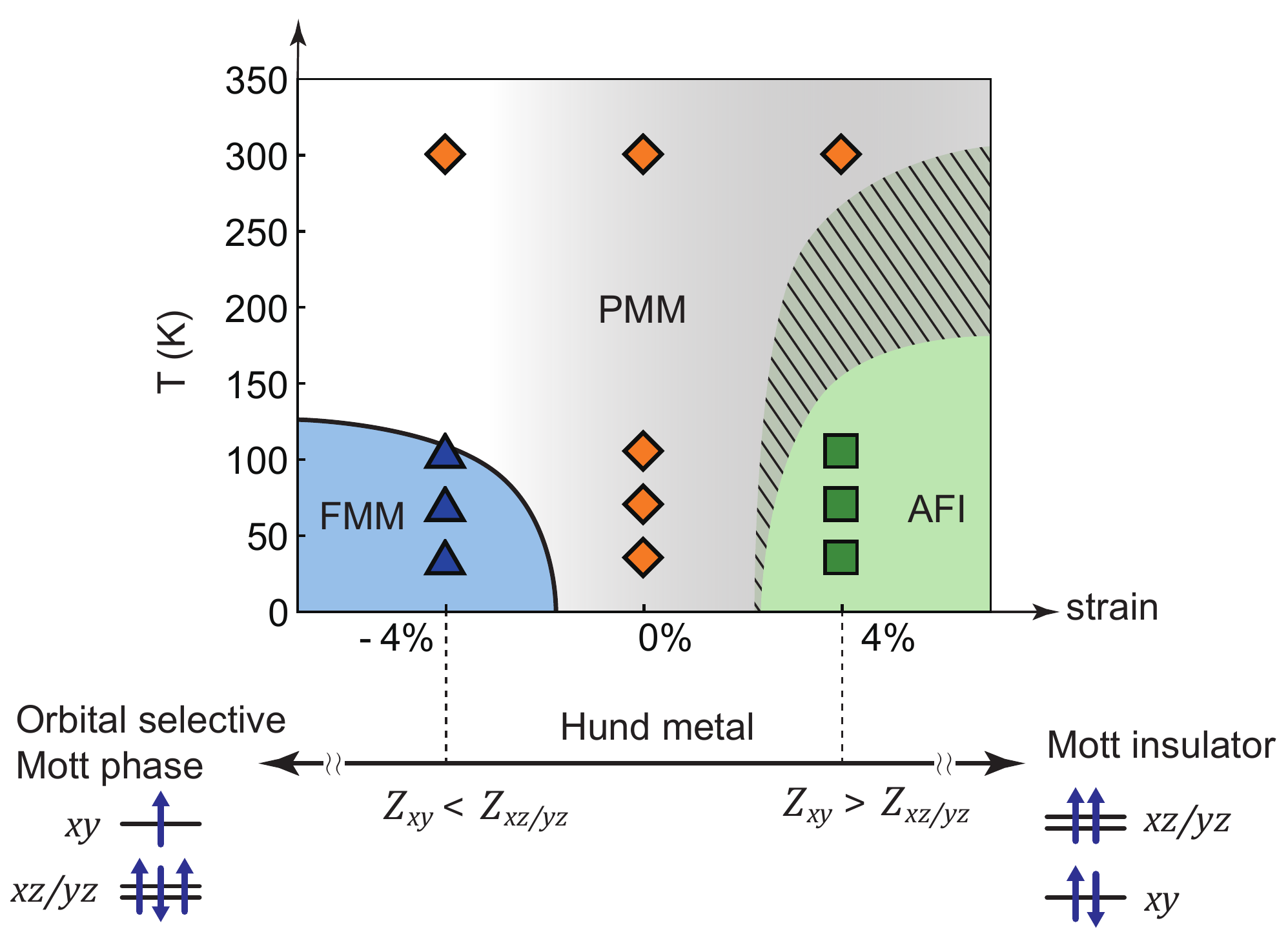}
\caption{
Temperature-strain dependent phase diagram of SRO-STO.
The DFT+DMFT results are denoted by blue triangle, orange rhombus, and green square for FM metallic phase (FMM), PM metallic phase (PMM), and AFM insulating (AFI) phase, respectively.
The gradient from gray to white color shows the variation of $Z_{xy}$ and $Z_{xz/yz}$ values, $Z_{xy}>Z_{xz/yz}$ (dark region) and $Z_{xy}<Z_{xz/yz}$ (light region).
The phase boundaries are drawn schematically. The hatched region between AFI and PMM phases indicates asserted first order phase transition.
Limiting cases are multi-orbital Mott insulating phase with half-filled (fully filled) $xz/yz$ ($xy$) orbital ($Z_{xy}>Z_{xz/yz}$), and orbital selective
Mott phase having the half-filled (3/2 filled) $xy$ ($xz/yz$) orbital ($Z_{xy}<Z_{xz/yz}$).
%Strain-temperature dependent phase diagram of SRO-STO.
%The gray color indicates the electronic structure
%with $Z_{xy}>Z_{xz/yz}$
%which is connected to the multi-orbital Mott insulating
%phase with half-filled (fully filled) $xz/yz$ ($xy$) orbital.
%The white color indicates the electronic structure
%with $Z_{xy}<Z_{xz/yz}$
%which is connected to the orbital selective
%Mott phase having the half-filled (3/2 filled) $xy$ ($xz/yz$) orbital.
%FM metallic phase (FMM), PM metallic phase (PMM), and AFM insulating (AFI) phase
%in the DFT+DMFT are denoted as blue triangle, orange rhombus,
%and green square, respectively.
%The FMM phase for the -4\% strain is covered with light blue color,
%and the AFI phase for the +4\% strain is covered with light green color.
%The shaded area above the AFI phase is
%for the co-existence of the PMM and AFI.
%The black line between FMM and PMM is a schematic phase boundary.
\label{fig:fig3}
}
\end{figure}
%%%%%%%%%%%%%%%%%%%%%%%%%%%%%%%%%%%%%%%%%%%%%%%%%%%%%%%%%%%%%%%%

{\it Phase diagram.}--
By collecting all the data on the electronic and magnetic properties, we finally construct a comprehensive strain-temperature phase diagram of the SRO-STO system (see Fig.~\ref{fig:fig3}).
It is illustrated that the strain-induced variation of the crystal field splitting is an essential physical parameter for emergent quantum phases.
In the tensile (compressive) strain, the $xz/yz$ ($xy$) orbital energy level is lifted up with respect to the $xy$ ($xz/yz$) orbital, driving the system toward the multi-orbital Mott insulator (orbital selective Mott phase).
This strain induced manipulation of physical regimes in SRO-STO
is related to the emergence of various magnetism.

Our phase diagram in Fig.~\ref{fig:fig3} introduces two accessible routes for the magnetic phase transition employing
the epitaxial strain engineering starting from the paramagnetic Hund metal phase of
the unstrained system.
First, by applying the compressive strain, one reaches the FM metal phase from
the competition with the SDW and the AFM phases. This is very interesting because,
while there are a few reports on the static magnetism upon doping or external perturbation~\cite{Carlo2012,grinenko2021split,Barden2002,BKim2022_2},
due to the fragile nature of FM, most of them are AFM or SDW and no FM has been stabilized for layered ruthenates.
%Our phase diagram introduces two accessible routes for the magnetic phase transition employing the epitaxial strain engineering. First, starting from the paramagnetic metal phase of the unstrained case, by applying the compressive strain, one can reach the FM metal phase from the competition with SDW phase. This is very interesting because, while there are a few reports on the static magnetism, due to the fragile nature of FM, most of them are AFM or SDW and no FM has been stabilized for layered ruthenates.
Together with orbital-selective characteristic of the compressive case, this phase can offer an unreported areas for further studies. Note that SrRuO$_3$ is also a FM metal, but the system is a three-dimensional one and has a much mild electronic correlation~\cite{Dang2015,MKim2015,Hahn2021}. When the tensile strain is applied, the other magnetic instabilities are quickly muted and a checkerboard AFM phase with a strong insulating electronic structure can be obtained. This phase is similar to the one from Ca$_2$RuO$_4$. As the temperature is increased, the system goes through the first-order type transition into the PM metal phase. We expect that there exists a coexistence phase of PM metal and AFM insulator based on the first-order type transition and unattainable convergence, which is denoted as the shaded area in Fig.~\ref{fig:fig3}. This is very different to the compressive strain case, where the FM transition is expected to be a second-order one.
The different mechanisms of magnetism can be a reason for
the contrasting transition, (i) the $itinerant$ Stoner mechanism induced FM emerges in the metallic phase
and (ii) the $local$ superexchange mechanism induced AFM emerges in the insulating phase.
We note that the AFM insulating phase for the tensile strain emerges
with the enhancement of the orbital polarization, having nearly integer filling of $n_{xz/yz}$=1.01 and $n_{xy}$=2.00, in comparison with the non-integer filling in the PM metallic phase of the tensile strain in Fig. \ref{fig:fig1}(d).
On the other hand, the FM metallic phase for the compressive strain
emerges without changes of the orbital occupation in comparison with the non-integer filling for the PM metallic
phase of the compressive strain in Fig. \ref{fig:fig1}(d).

{\it Conclusion.}-- We have investigated the strain-temperature dependent electronic and magnetic properties of the SRO-STO heterostructure system within the DFT+DMFT method. When unstrained, the system has a slightly enhanced magnetic tendency with a similar degree of electronic correlation over bulk layered counterparts. But, such a small difference enables a plethora of interesting phases when combined with the workable epitaxial strain. We presented the strain-temperature phase diagram of the SRO-STO system, which can access various electronic and magnetic phases observed from the diverse bulk layered system as well as the unreported magnetic and electronic phases such as FM metal with orbital-selectiveness.
We expect our work can guide the future experimental and theoretical directions towards the engineering of the layered ruthenates, and other correlated systems with competing physical energy scales.

\section{acknowledgements}
The authors like to acknowledge the support from Advanced Study Group program from PCS-IBS.
MK was supported by KIAS Individual Grants(CG083501).
CJK was supported by the NRF grant (No. 2022R1C1C1008200) and KISTI Supercomputing Center (Project No. KSC-2021-CRE-0580).
BK acknowledges support by NRF grand No. 2021R1C1C1007017, 2022M3H4A1A04074153, and KISTI supercomputing Center (Project No. KSC-2021-CRE-0605).
JHH acknowledge financial support from the Institute for Basic Science in the Republic of Korea through the project IBS-R024-D1.
KK is supported by KAERI internal R\&D Program (No. 524460-22).
Part of the calculation is supported by the CAC at KIAS.

\bibliography{refs_SRO-STO-Phase_Diagram}

%%%%% Supplemental Material %%%%

\renewcommand{\thetable}{S\arabic{table}}
\renewcommand{\thefigure}{S\arabic{figure}}
\renewcommand{\thetable}{S\arabic{table}}
\renewcommand\theequation{S\arabic{equation}}
\setcounter{table}{0}
\setcounter{figure}{0}
\setcounter{equation}{0}
\renewcommand{\bibnumfmt}[1]{[S#1]}
\renewcommand{\citenumfont}[1]{S#1}
%\setcounter{biburllcpenalty}{0}

%useful definitions here
\def\cred{\color{red}}
\def\cblue{\color{blue}}

\onecolumngrid

\clearpage

\begin{center}
{\bf \large
{\it Supplemental Material:}\\
Tunable electronic and magnetic phases in layered ruthenates: \\
SrRuO$_{3}$-SrTiO$_{3}$ heterostructure upon strain
}

\vspace{0.2 cm}
Minjae Kim$^{1}$, Chang-Jong Kang$^{2}$, Jae-Ho Han$^{3}$, Kyoo Kim$^{4}$, and Bongjae Kim$^{5}$
\vspace{0.1 cm}

{\small{\it
$^{1}$Korea Institute for Advanced Study, Seoul 02455, South Korea \\
$^{2}$Department of Physics, Chungnam National University, Daejeon 34134, South Korea \\
$^{3}$Center for Theoretical Physics of Complex Systems,
Institute for Basic Science (IBS), Daejeon 34126, South Korea \\
$^{4}$Korea Atomic Energy Research Institute (KAERI), Daejeon 34057, South Korea \\
$^{5}$Department of Physics, Kunsan National University, Gunsan 54150, South Korea
}
}
\end{center}

%%%%% Supplemental Material %%%%

\section{Method}

In this section, we describe the detailed computational methods for our DFT+DMFT calculations on SRO-STO.

\subsection{crystal structure}
For the crystal structure of the SRO-STO heterostructure, we have adapted the optimized structures in the DFT framework from Ref.~\cite{kim2020srruo}. We assumed the in-plane lattice parameter of SrTiO$_{3}$, 3.905 {\AA}, as a reference for unstrained case (0\%), and biaxial strain of -4\% and +4\% cases were considered by fixing in-plane lattice parameters. The octahedral rotation ($\sqrt{2} \times \sqrt{2}$) is allowed
($a^{0}a^{0}c^{-}$ Glazer tilting of octahedra) in the tetragonal unit cell. Refer to Ref.~\cite{kim2020srruo} for the details.

\subsection{$t_{2g}$ only projector method of DFT+DMFT}

%Here, we describe the low energy $t_{2g}$ only projector method for the DFT+DMFT framework.
%This method have advantages for the computation of low temperature electronic structures and magnetic phases due to the lower
%computational cost.

Electronic structure calculation in the DFT+DMFT framework~\cite{georges1996dynamical,kotliar2006electronic}
from the $t_{2g}$ only projector method were performed using the full potential implementation in the TRIQS library~\cite{aichhorn2016triqs,parcollet2015triqs}.
The DFT part of the computations in the local density approximation were performed employing the WIEN2k package~\cite{blaha2020wien2k}.
We used 16 $\times$ 16 $\times$ 11 $k$-point mesh for the Brillouin zone integration.
Wannier-like $t_{2g}$ orbitals were constructed from the Kohn-Sham bands, which includes 6 $t_{2g}$ bands and 7 topmost $O(2p)$ bands.
Energy ranges for related Kohn-Sham bands at $k = \Gamma$ are
[-2.7, 0.6] eV, [-2.5, 0.6] eV, and [-2.2, 0.6] eV, for -4\%, 0\%, and +4\% strains, respectively.
We used the full rotationally invariant Kanamori interaction with $U$ = 2.3 eV and $J$ = 0.4 eV.
This setup successfully described the physical properties of the ruthenates in previous studies~\cite{ricco2018situ,sutter2017hallmarks,sutter2019orbitally,Mravlje2011,Dang2015,Stricker2014}.
The quantum impurity problem in the DMFT was solved using the continuous-time hybridization-expansion quantum Monte Carlo impurity solver as implemented in the TRIQS library~\cite{seth2016triqs,gull2011continuous}. We perform the charge self-consistent DFT+DMFT computation for the PM phase for each strains with the convergence criteria of the total energy of 0.002 (Ry/formula unit).
For the computation of magnetic phases, the converged charge densities from the PM calculations are fixed, and we allowed the spin symmetry breaking in the local Greens function and the self-energy of the $t_{2g}$ orbital according to the magnetic orders. When stabilizing the magnetic ground state, we first started from the converged local Green's function and the self-energy of the $t_{2g}$ orbital from PM phase calculations, then we applied local magnetic field (in the order of $\sim$ 5 meV), homogeneous one for the FM and staggered one for the AFM, to the local Hamiltonian for the initial a few DMFT iterations. We checked that stable magnetic phases in the present result are well-converged and robust for the DMFT iteration in the absence of the local magnetic field.

\subsection{$e_{g}$+$t_{2g}$ projector method of DFT+DMFT}

%Here, we describe the $e_{g}$+$t_{2g}$ projector method within the DFT+DMFT framework.
%This method have advantages for the confirmation of the important Ru $t_{2g}$ bands position with respect to Ru $e_{g}$ and O $(2p)$ bands,
%while the computation is more expensive than the $t_{2g}$ only projector method.

In the Ru $e_{g}$+$t_{2g}$ projector scheme, we performed fully charge self-consistent
DFT+DMFT calculations implemented in the all-electron full-potential WIEN2k package~\cite{blaha2020wien2k}
with formalisms described in Ref.~\cite{Haule2010}.
We choose a wide hybridization energy window from -10 eV to 10 eV with respect to the Fermi level $E_{\text{F}}$.
All the five Ru-$4d$ ($e_{g}$+$t_{2g}$) orbitals are considered as correlated ones and
the fully rotational invariant form is applied for a local Coulomb interaction Hamiltonian with the on-site Coulomb repulsion $U=4.5$ eV and Hund's coupling $J_{H}=1$ eV. The Coulomb parameters $U$ and $J_{H}$ are confirmed by previous DFT+DMFT calculations
on several metallic ruthenates in the $e_{g}$+$t_{2g}$ projector method~\cite{Deng2016}.
The continuous time quantum Monte Carlo (CTQMC)~\cite{Werner2006,Haule2007}
is adopted for a local impurity solver. We use a generalized gradient approximation (GGA)~\cite{Perdew1996} for the exchange-correlation functional and subtracted the nominal double counting term.
In the charge self-consistent calculations, 19 $\times$ 19 $\times$ 13 $k$-point mesh is used for the Brillouin zone integration.
For dynamical magnetic response function calculations, we computed both the one-particle Green's function and the two-particle vertex function
with formalisms described in Ref.~\cite{Park2011}
within the DFT+DMFT method.

\subsection{Density of states for the correspondence of $t_{2g}$ only projector method and $e_{g}$+$t_{2g}$ projector method}

%%%%%%%%%%%%%%%%%%%%%%%%%%%%%%%%%%%%%%%%%%%%%%%%%%%%%%%%%%%%%%%%
\begin{figure}[t]
\includegraphics[width=\columnwidth]{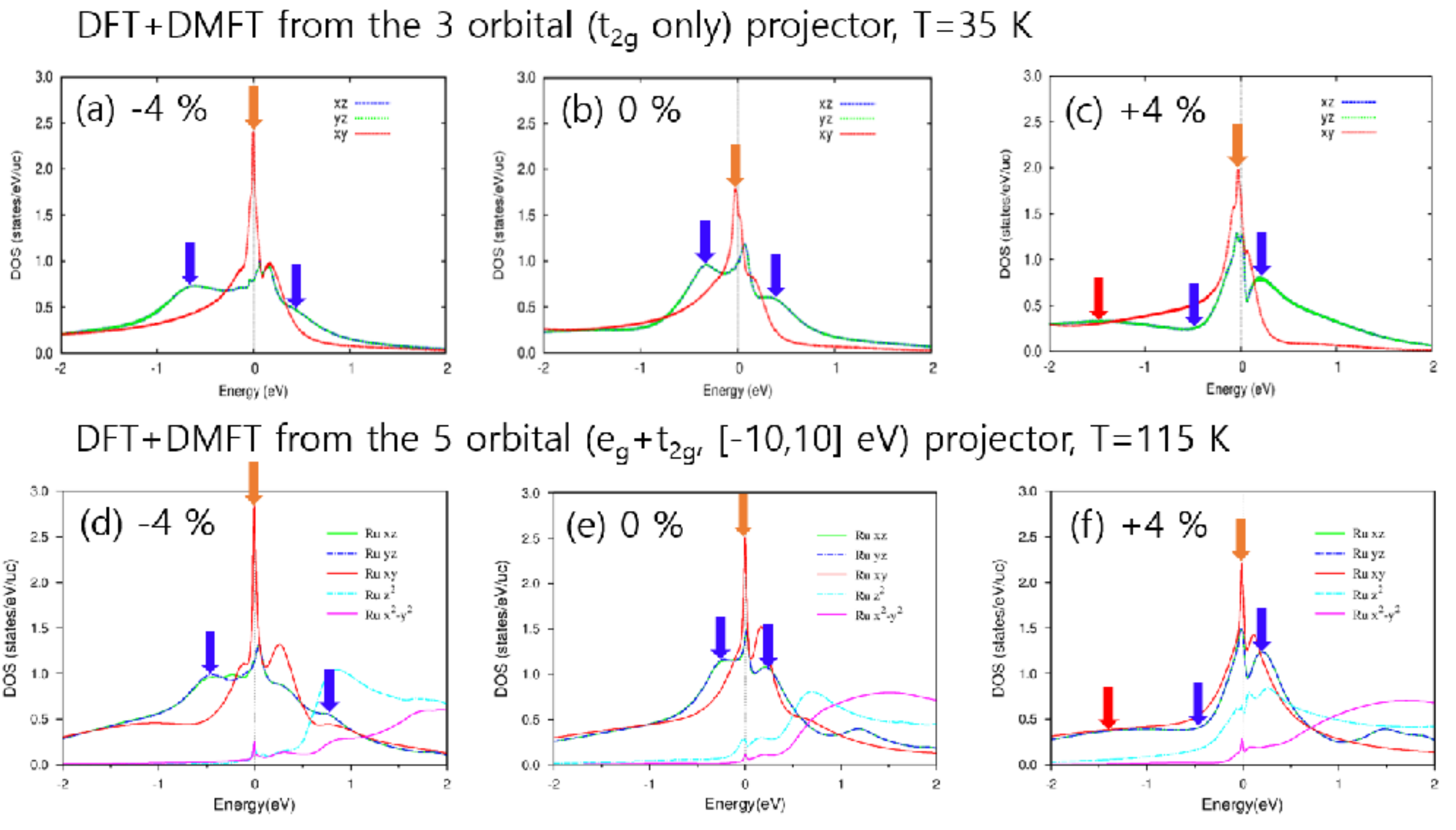}
\caption{
(a), (b), and (c) show density of states
from the $t_{2g}$ only projector method for the DFT+DMFT.
Strains of (a) -4\%, (b) 0\%, and (c) +4\% are considered.
PM state is forced at 35 K.
(d), (e), and (f) show density of states
from the $e_{g}$+$t_{2g}$ projector method for the DFT+DMFT.
Strains of (d) -4\%, (e) 0\%, and (f) +4\% are considered.
PM state is forced at 115 K.
The orange arrows indicate the evolution of the van-Hove singularity peak of the $xy$ orbital.
The blue arrows indicate the evolution of the hump of the $xz/yz$ orbital
near the Fermi level.
The red arrow indicates the hump for the binding energy of $(U+J)/2 \sim 1.35$ eV only for the +4\% strain, dominantly from $xz/yz$ orbitals.
\label{fig:DOS}
}
\end{figure}
%%%%%%%%%%%%%%%%%%%%%%%%%%%%%%%%%%%%%%%%%%%%%%%%%%%%%%%%%%%%%%%%
%%%%%%%%%%%%%%%%%%%%%%%%%%%%%%%%%%%%%%%%%%%%%%%%%%%%%%%%%%%%%%%%
\begin{figure}[t]
\includegraphics[width=8cm]{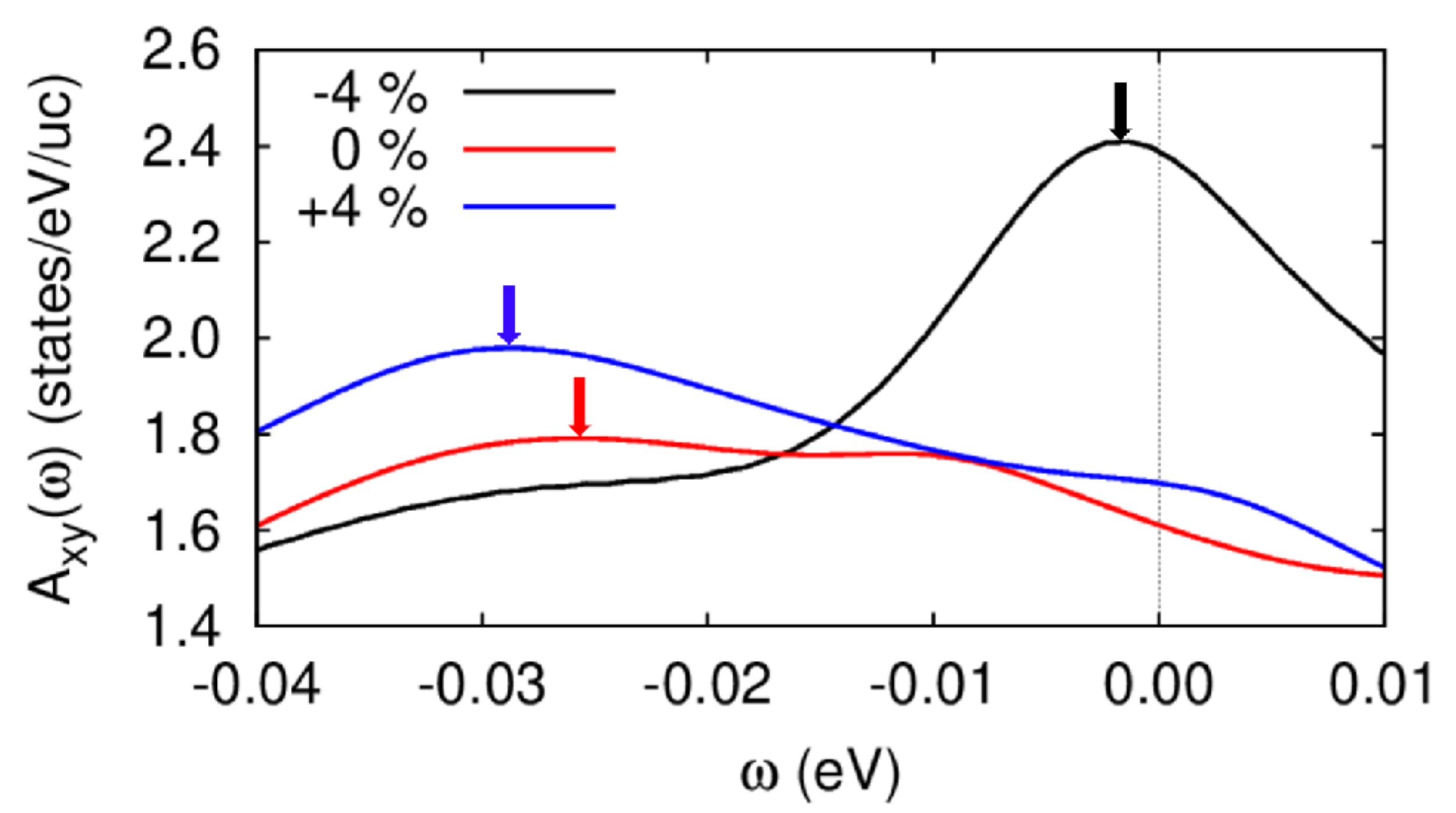}
\caption{The blowup figures for the density of states
of the $xy$ orbital ($A_{xy}(\omega)$) near the Fermi level,
for -4 \% (black), 0 \% (red), and +4 \% (blue) strains.
The van-Hove singularity peaks are indicated with arrows for each strains.
The result from the $t_{2g}$ only projector scheme is shown.
\label{fig:vHs}
}
\end{figure}
%%%%%%%%%%%%%%%%%%%%%%%%%%%%%%%%%%%%%%%%%%%%%%%%%%%%%%%%%%%%%%%%

In Fig.~\ref{fig:DOS}, we compare the DFT+DMFT spectral functions $A(\omega)$ of the PM SRO-STO system from two different projector scheme. We confirm the overall agreement of the key features:
(i) significant increase of the Ru $xy$-orbital peak near the Fermi level upon the compressive strain;
(ii) hump features from Ru $xz/yz$ orbitals near the Fermi level (blue arrows in Fig.~\ref{fig:DOS});
(iii) broad local maximum for +4 \% tensile strain (red arrow) at around a binding energy of $(U+J)/2\sim1.35$ eV as an indication of Mott regime. We assert here that the two project approaches match each other.
In Fig.~\ref{fig:vHs}, we present the $A(\omega)$ of the $xy$ orbital ($A_{xy}(\omega)$)
near the Fermi level, in the PM SRO-STO system from the $t_{2g}$ projector scheme.
It is clearly shown that the van-Hove singularity peak moves towards
the Fermi level upon the compressive strain due to the increased $xy$ orbital
energy level as shown in Table~\ref{table:orbital}.
The significant increasement of the $A_{xy}(0)$ for the compressive strain
is consistent with the density of states in the FM phase for the compressive strain in the main text,
having the dominant spin splitting in the $xy$ orbital.

\section{Hund Metallicity for unstrained SRO-STO}

%%%%%%%%%%%%%%%%%%%%%%%%%%%%%%%%%%%%%%%%%%%%%%%%%%%%%%%%%%%%%%%%
\begin{figure}[t]
\includegraphics[width=14cm]{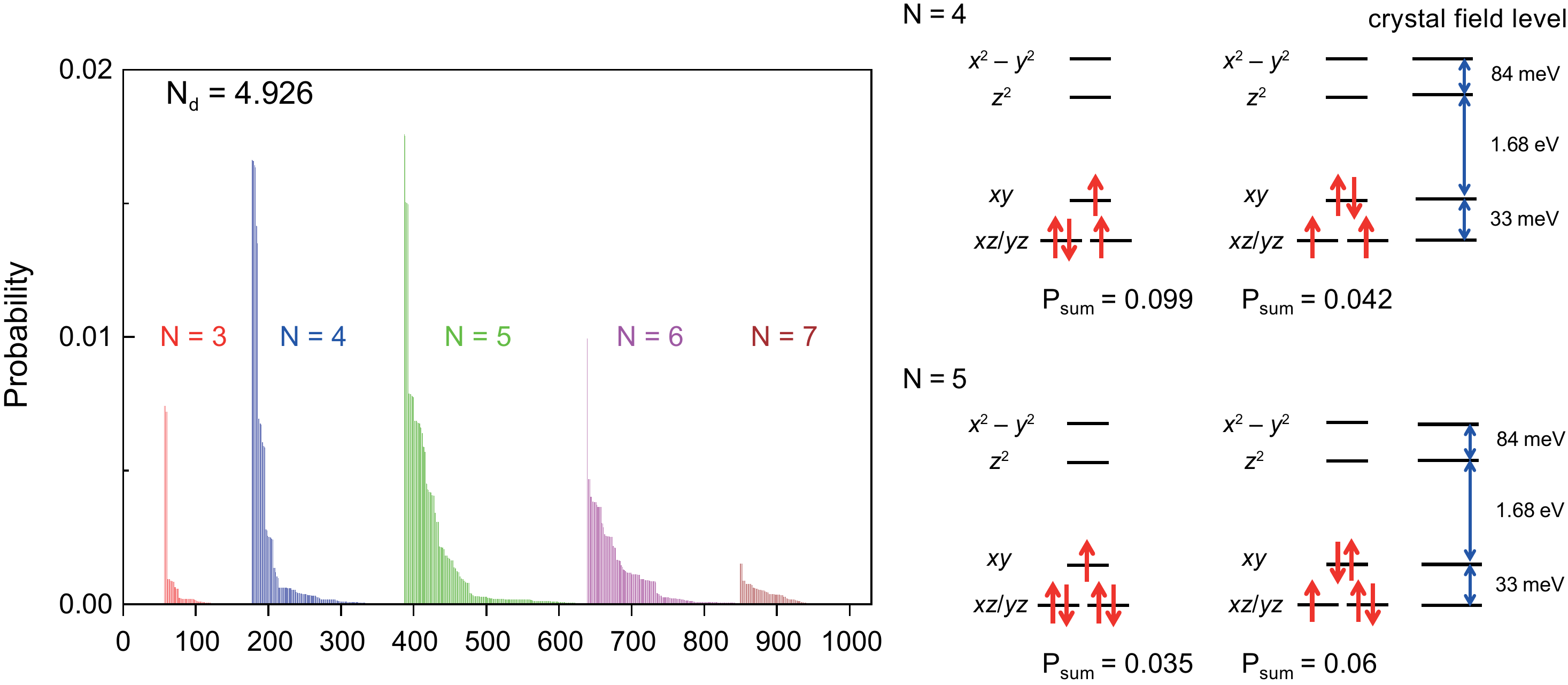}
\caption{Histogram of multiplets
for the un-strained SRO-STO in the DFT+DMFT
computation. The $e_{g}$+$t_{2g}$ projector method
is used. The most and the second most dominant
electronic configuration for the occupancies $N$ = 4 and 5
are shown.
The crystal field energy level for Ru($d$)
orbital in the $e_{g}$+$t_{2g}$ projector method
is shown.
P$_{sum}$ is the probability sum of the histogram
for the multiplet,
considering the spin and $xz/yz$ orbital degeneracies.
\label{fig:histogram}
}
\end{figure}
%%%%%%%%%%%%%%%%%%%%%%%%%%%%%%%%%%%%%%%%%%%%%%%%%%%%%%%%%%%%%%%%
%%%%%%%%%%%%%%%%%%%%%%%%%%%%%%%%%%%%%%%%%%%%%%%%%%%%%%%%%%%%%%%%
\begin{figure}[t]
\includegraphics[width=\columnwidth]{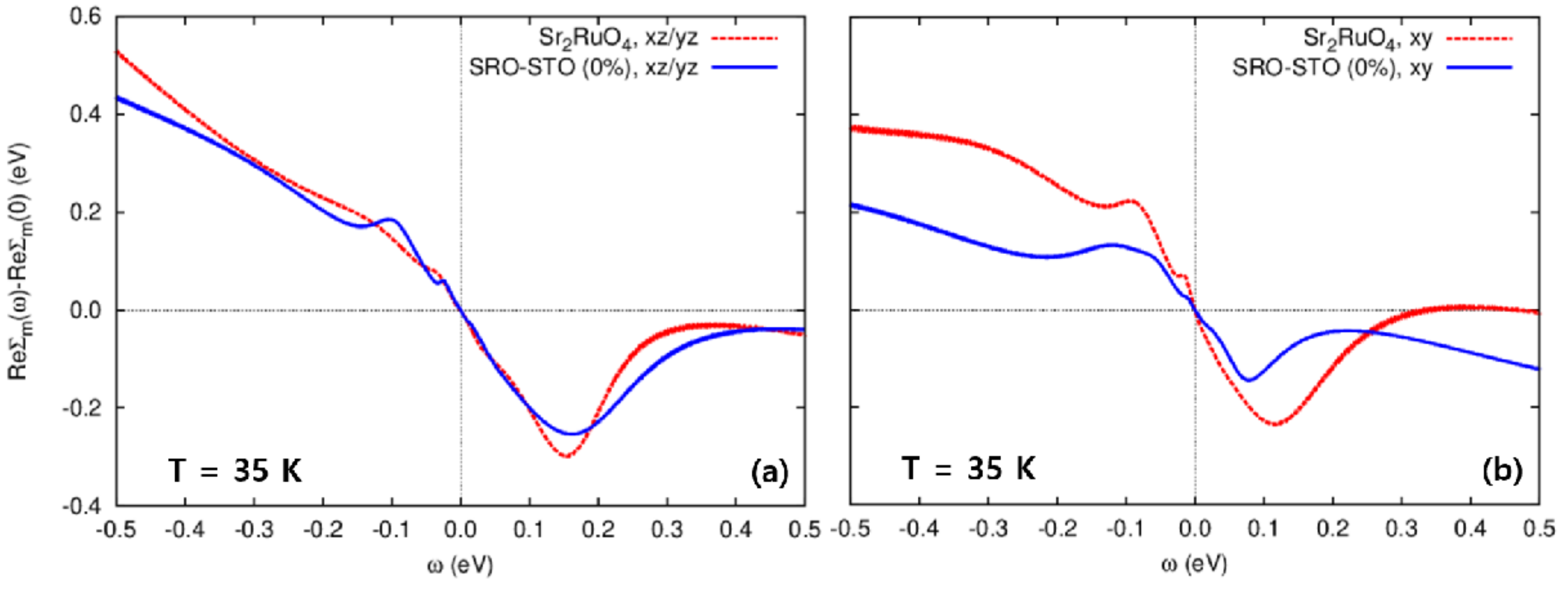}
\caption{Real part self-energies
in the real frequency axis
for SRO-STO with 0\% strain and Sr$_{2}$RuO$_{4}$
at the temperature of 35 K.
The left panel (a) is for the $xz/yz$ orbital
and the right panel (b) is for the $xy$ orbital, respectively.
\label{fig:214_self_energy}
}
\end{figure}
%%%%%%%%%%%%%%%%%%%%%%%%%%%%%%%%%%%%%%%%%%%%%%%%%%%%%%%%%%%%%%%%

In this section, we show the Hund metal characteristics of the unstrained paramagnetic SRO-STO system.
Fig.~\ref{fig:histogram} presents the DMFT valence histogram computed with the Ru $e_{g}$+$t_{2g}$ projector scheme method.
We find sizable probabilities over several different electronic configurations with $N$ = 4, 5, and 6, indicating the strong charge fluctuation. Note that in a Mott system, the charge fluctuation is substantially blocked by the electronic Coulomb interaction and, therefore,
certain atomic multiplet dominates the DMFT valence histogram and nearly integer occupancy is realized, which is not the case for our system.

In the $N$ = 4 sector, high-spin atomic multiplets have significant probabilities, thereby presenting that the system is under the influence of strong Hund's coupling. Besides, in the $N$ = 5 sector, the most probable state is derived from adding one electron to most probable state for $N$ = 4.
Strong Hund's coupling induced high-spin atomic multiplets, a typical hallmark of Hund metal~\cite{haule2009coherence}, can be identified here.

In Fig.~\ref{fig:214_self_energy}, we compare the real part of the self-energy in the real frequency $\omega$ of unstrained SRO-STO to Sr$_{2}$RuO$_{4}$, a prototypical Hund metal~\cite{Fabian2020,Mravlje2011}. One can see the Fermi-liquid-like self-energy in a narrow $\omega$ range at around the Fermi level as well as the self-energy humps at around $\omega = \pm0.1$ eV for both systems. The quasiparticle residues $Z$ of unstrained SRO-STO are 0.30 ($xz/yz$) and 0.35 ($xy$), which are not far from the values from Sr$_{2}$RuO$_{4}$, 0.30 ($xz/yz$) and 0.20 ($xy$).

\section{Strain and orbital selective electronic correlation}

%%%%%%%%%%%%%%%%%%%%%%%%%%%%%%%%%%%%%%%%%%%%%%%%%%%%%%%%%%%%%%%%
\begin{figure}[t]
\includegraphics[width=\columnwidth]{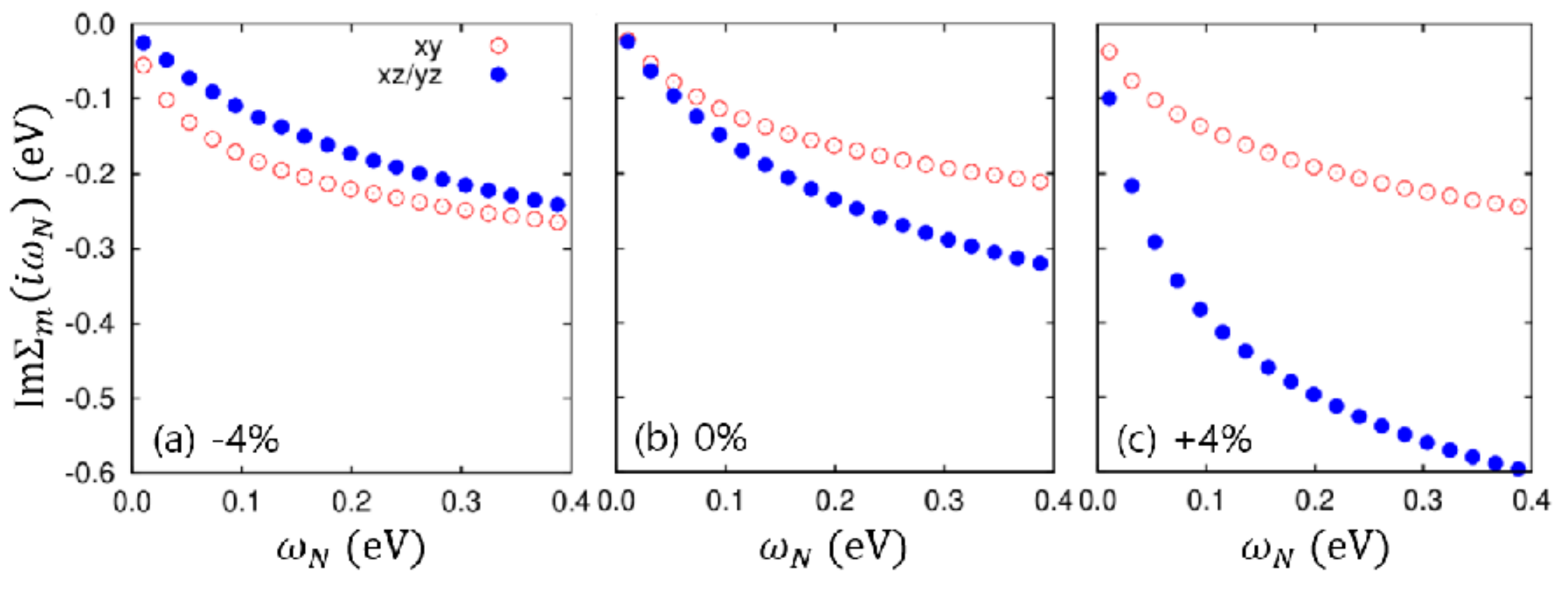}
\caption{Strain dependent imaginary part self-energy
in the Matsubara frequency for (a) -4\%, (b) 0\%, and (c) +4\%.
PM constraint is applied with the temperature of 35 K.
\label{fig:self_energy}
}
\end{figure}
%%%%%%%%%%%%%%%%%%%%%%%%%%%%%%%%%%%%%%%%%%%%%%%%%%%%%%%%%%%%%%%%
%%%%%%%%%%%%%%%%%%%%%%%%%%%%%%%%%%%%%%%%%%%%%%%%%%%%%%%%%%%%%%%%%%
%\begin{table}[t]
%\centering
%\caption{Strain dependent (i) the in-plane Ru-O bond length $d_{Ru-O(x)}$,
%(ii) the out-of-plane Ru-O bond length $d_{Ru-O(z)}$,
%(iii) the local crystal
%field in the $t_{2g}$ manifold ($\epsilon_{xz/yz}$-$\epsilon_{xy}$),
%(iv) the occupancy in the $t_{2g}$ manifold ($n_{xy}$ and $n_{xz/yz}$),
%and (v) the quasiparticle residue in the $t_{2g}$ manifold ($Z_{xy}$ and $Z_{xz/yz}$), are presented.
%For all computations, PM state is forced at 35 K.}
%\begin{ruledtabular}
%\begin{tabular}{ || c || c | c | c ||}
%~ & ~-4 \% & ~0 \% & +4 \%  \\
%\hline \hline
%$d_{Ru-O(x)}$ (${\AA}$) & 1.91 & 1.97 & 2.05 \\
%\hline
%$d_{Ru-O(z)}$ (${\AA}$) & 2.05 & 2.00 & 1.94 \\
%\hline
%$\epsilon_{xz/yz}$-$\epsilon_{xy}$ (eV) & 0.300 & 0.549 & 0.592 \\
%\hline
%$n_{xy}$ & 1.46 & 1.59 & 1.70 \\
%\hline
%$n_{xz/yz}$ & 1.38 & 1.32 & 1.20 \\
%\hline
%$Z_{xy}$ & 0.21 & 0.35 & 0.37 \\
%\hline
%$Z_{xz/yz}$ & 0.43 & 0.30 & 0.11 \\
%\end{tabular}
%\end{ruledtabular}
%\label{table:orbital}
%\end{table}
%%%%%%%%%%%%%%%%%%%%%%%%%%%%%%%%%%%%%%%%%%%%%%%%%%%%%%%%%%%%%%%%%

%%%%%%%%%%%%%%%%%%%%%%%%%%%%%%%%%%%%%%%%%%%%%%%%%%%%%%%%%%%%%%%%
\begin{table}
\caption{Strain dependent (i) the in-plane Ru-O bond length $d_{\textrm{Ru-O}(x)}$,
(ii) the out-of-plane Ru-O bond length $d_{\textrm{Ru-O}(z)}$,
(iii) the local crystal
field in the $t_{2g}$ manifold ($\epsilon_{xz/yz}-\epsilon_{xy}$),
(iv) the occupancy in the $t_{2g}$ manifold ($n_{xy}$ and $n_{xz/yz}$),
and (v) the quasiparticle residue in the $t_{2g}$ manifold ($Z_{xy}$ and $Z_{xz/yz}$), are presented.
For all computations, PM state is forced at 35 K.}
\begin{ruledtabular}
\begin{tabular}{lddd}
& \multicolumn{1}{c}{\textrm{\ -4 \%}} & \multicolumn{1}{c}{\textrm{\ 0 \%}} & \multicolumn{1}{c}{\textrm{\ +4 \%}}  \\
\hline
(i) $d_{\textrm{Ru-O}(x)}$ (\AA) & 1.91 & 1.97 & 2.05 \\
(ii) $d_{\textrm{Ru-O}(z)}$ (\AA) & 2.05 & 2.00 & 1.94 \\
(iii) $\epsilon_{xz/yz} - \epsilon_{xy}$ (eV) & 0.300 & 0.549 & 0.592 \\
(iv) $n_{xy}$ & 1.46 & 1.59 & 1.70 \\
\hspace{14pt} $n_{xz/yz}$ & 1.38 & 1.32 & 1.20 \\
(v) $Z_{xy}$ & 0.21 & 0.35 & 0.37 \\
\hspace{13pt} $Z_{xz/yz}$ & 0.43 & 0.30 & 0.11 \\
\end{tabular}
\end{ruledtabular}
\label{table:orbital}
\end{table}
%%%%%%%%%%%%%%%%%%%%%%%%%%%%%%%%%%%%%%%%%%%%%%%%%%%%%%%%%%%%%%%%

In Fig.~\ref{fig:self_energy}, we present the self-energy of $xz/yz$ and $xy$ orbitals of SRO-STO for different strains. Along with our quasiparticle residues $Z$ data (see Table~\ref{table:orbital}), the orbital-selective characteristics for strained cases are well-captured. For 0\% strain, the $xz/yz$ and $xy$ orbitals have a relatively similar degree of electronic correlation and have similar quasiparticle residues $Z$. When the compressive strain of -4\% is applied, stronger electronic correlation in the $xy$ orbital than the $xz/yz$ orbital can be seen and \emph{vice versa} for the tensile strain cases.

\section{Fermi Surface}

%%%%%%%%%%%%%%%%%%%%%%%%%%%%%%%%%%%%%%%%%%%%%%%%%%%%%%%%%%%%%%%%
\begin{figure}[t]
\includegraphics[width=\columnwidth]{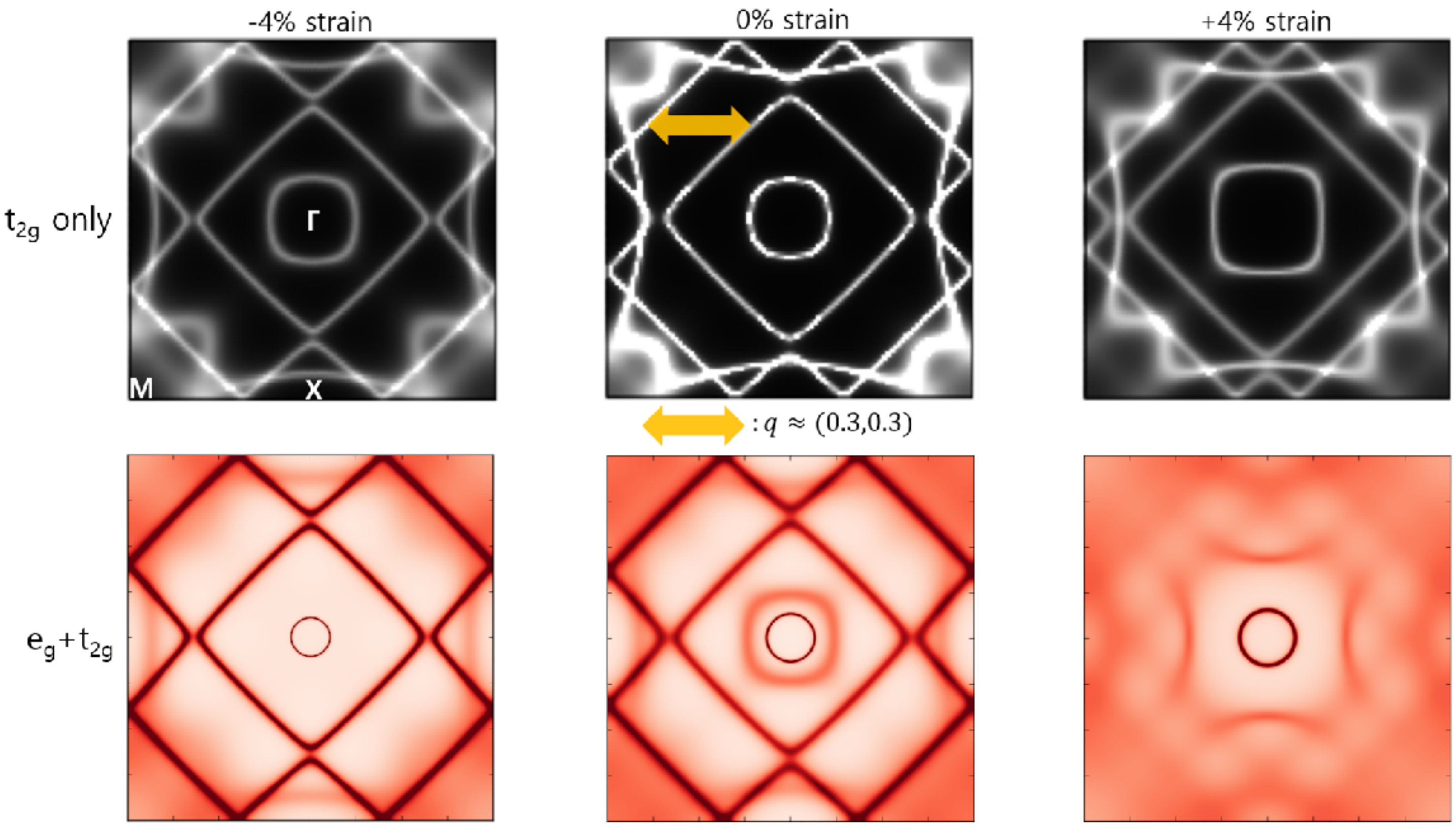}
\caption{Strain dependent Fermi surface
for -4\%, 0\%, and +4\%.
Upper panels are for $t_{2g}$ only projector method for the DFT+DMFT.
Fermi surface nesting vector
for the SDW ($q \sim (0.3,0.3)$)
is denoted from yellow arrow.
Lower panels are for the $e_{g}$+$t_{2g}$ projector method for the DFT+DMFT.
For the $t_{2g}$ only projector method, the PM state is forced at 35 K.
For the $e_{g}$+$t_{2g}$ projector method, the PM state is forced at 115K.
\label{fig:FS}
}
\end{figure}
%%%%%%%%%%%%%%%%%%%%%%%%%%%%%%%%%%%%%%%%%%%%%%%%%%%%%%%%%%%%%%%%

In Fig.~\ref{fig:FS}, we present the strain-dependent Fermi surfaces of SRO-STO systems both from the $t_{2g}$ only and the $e_{g}$+$t_{2g}$ projector scheme. As with other electronic features, both methods give a reasonably consistent result.
For example, there is a grim Fermi pocket at around $M$ for 0\% and -4\% strains, which originates from the van-Hove singularity in the Ru $xy$ orbital.
Also, the compressive strain makes the Fermi surface nesting with $q_{\text{SDW}} \sim (0.3,0.3)$ slightly stronger, while the tensile strain makes it weaker. This is consistent with our dynamical structure factor calculations in the main text.
Still, we found that there are minor differences between two approaches.
For example, there exist additional electron pocket, originated from Ru $e_{g}$, at $\Gamma$ for five-orbital case.
However, our main discussion is not affected by the minor details of the Fermi surface morphology.

\section{Dynamical spin structure factor}

%%%%%%%%%%%%%%%%%%%%%%%%%%%%%%%%%%%%%%%%%%%%%%%%%%%%%%%%%%%%%%%%
\begin{figure}[t]
\includegraphics[width=\columnwidth]{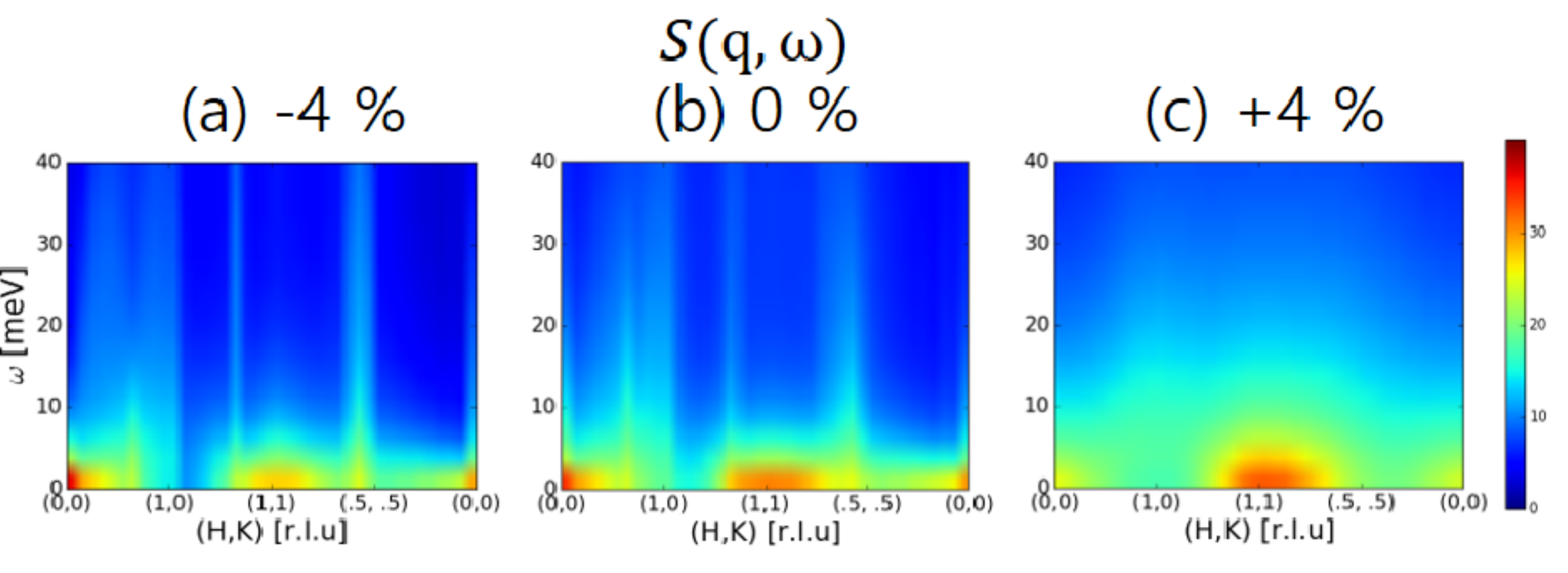}
\caption{
(a), (b), and (c) show momentum and frequency dependent dynamical spin structure factor from
(a) -4\% , (b) 0\%, and (c) +4\% strains, respectively, within L = 0 plane.
For all computations, PM state is forced at 115 K.
The $e_{g}+t_{2g}$ projector method is used.
\label{fig:S_factor1}
}
\end{figure}
%%%%%%%%%%%%%%%%%%%%%%%%%%%%%%%%%%%%%%%%%%%%%%%%%%%%%%%%%%%%%%%%
%%%%%%%%%%%%%%%%%%%%%%%%%%%%%%%%%%%%%%%%%%%%%%%%%%%%%%%%%%%%%%%%
\begin{figure}[t]
\includegraphics[width=14cm]{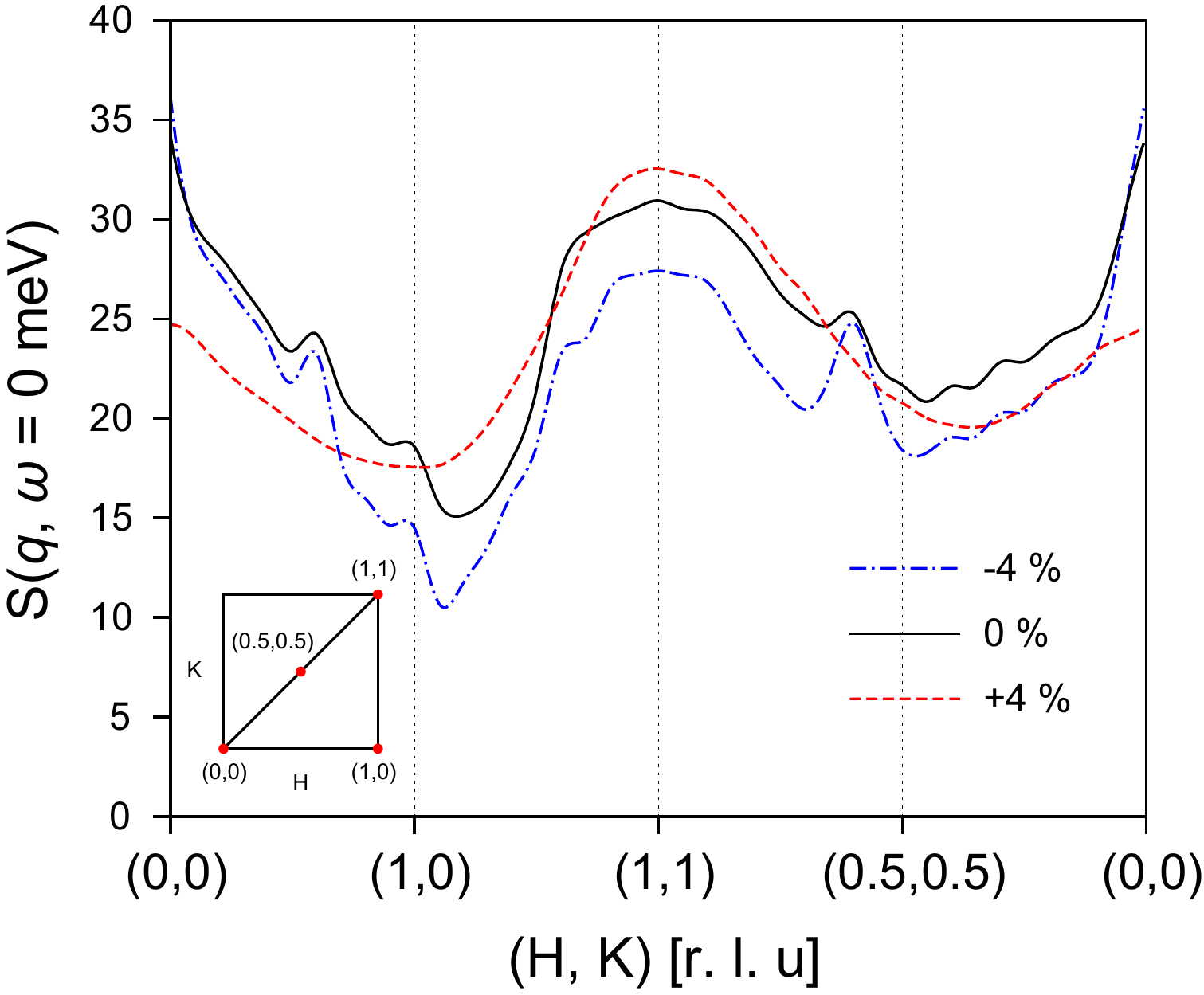}
\caption{Momentum dependent dynamical spin structure factor
at $\omega$ = 0 meV, within L = 0 plane for -4\%, 0\%, and +4\% strains.
For all computations, PM state is forced at 115 K.
The $e_{g}$+$t_{2g}$ projector method is used.
\label{fig:S_factor2}
}
\end{figure}
%%%%%%%%%%%%%%%%%%%%%%%%%%%%%%%%%%%%%%%%%%%%%%%%%%%%%%%%%%%%%%%%

We show the momentum $q$ and frequency $\omega$-dependent dynamical spin structure factor $S(q,\omega)$ of the SRO-STO system upon strains in Fig.~\ref{fig:S_factor1}.
The $q$ dependent $S(q,\omega)$ at $\omega$ = 0 meV is also plotted in Fig.~\ref{fig:S_factor2}.
We can directly confirm the strongest FM instability for -4\% strain, strong competition of FM, AFM, and SDW ones for unstrained one, and dominating AFM instability for 4\% strain cases.

\section{Temperature dependent evolution of magnetism}

%%%%%%%%%%%%%%%%%%%%%%%%%%%%%%%%%%%%%%%%%%%%%%%%%%%%%%%%%%%%%%%%
\begin{figure}[t]
\includegraphics[width=14cm]{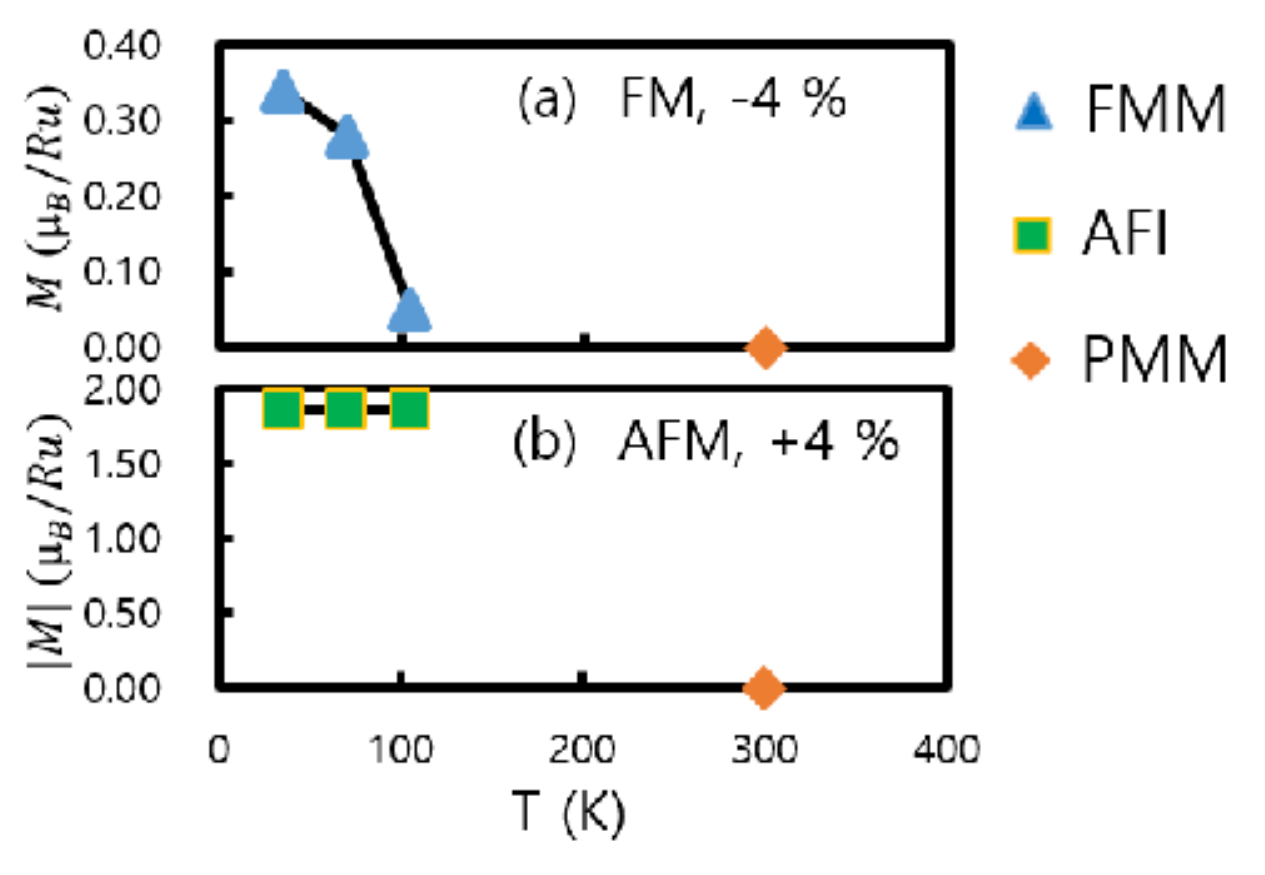}
\caption{(a) Temperature dependent evolution
of the magnetic momentum per Ru($t_{2g}$) for the
FM in the -4\% strain.
(b) Temperature dependent evolution
of the absolute magnetic momentum per Ru($t_{2g}$)
for the sublattice of the checkerboard AFM in the +4\% strain.
The $t_{2g}$ only projector method for the DFT+DMFT is used.
FM metal (FMM), AFM insulator (AFI), and PM metal (PMM)
phases obtained from the DFT+DMFT ($t_{2g}$ only projector method) are presented.
\label{fig:T_magnetism}
}
\end{figure}
%%%%%%%%%%%%%%%%%%%%%%%%%%%%%%%%%%%%%%%%%%%%%%%%%%%%%%%%%%%%%%%%

Here, we compare two distinct temperature evolution of FM and AFM orders for compressive and tensile strain cases. In Fig.~\ref{fig:T_magnetism}(a), we present the temperature-dependent magnetization $M$ for the FM phase under the compressive strain of -4\%. We find that the Curie temperature $T_{C}$ is about 100 K with continuous evolution of the moment size, indicating the second-order type transition. The saturated magnetization (at low temperature) is around 0.35 $\mu_{B}$. But for tensile strain case, the behavior is different. As shown in Fig.~\ref{fig:T_magnetism}(b), for the checkerboard AFM phase of tensile case, the local magnetic moment size of 1.87 $\mu_{B}$ does not show temperature dependence up to the 105 K. The magnetic moment is from the nearly fully-occupied $xy$ and half-filled $xz/yz$ orbitals for Ru atom. For temperature above 300 K, the AFM insulating phase is unstable and the PM metallic phase is found. In between, say at intermediate temperature of 200 K, the DFT+DMFT calculations does not converge into both the AFM and the PM.
This inaccessible convergence suggests a possible coexistence phase of the AFM insulator and the PM metal having the first-order type transition.

\bibliography{refs_SRO-STO-Phase_Diagram}

\end{document}